\documentclass[preprint,12pt]{elsarticle}

\usepackage{amssymb}

\usepackage{graphicx,color}
\usepackage[usenames,dvipsnames]{xcolor}
\usepackage[section]{placeins}
\usepackage[normalem]{ulem}

\newcommand{\cd}{\makebox[0.08cm]{$\cdot$}}

\journal{Journal of Subatomic Particles and Cosmology}

\usepackage{amssymb}
\usepackage{amsmath}

\usepackage{epsfig,color,graphicx}
\usepackage{xcolor}
\usepackage{subfigure}

\colorlet{darkred}{red!80!black}
\colorlet{darkgreen}{green!50!black}
\colorlet{darkblue}{blue!50!black}

\usepackage{soul}

\usepackage[pdftex,hypertexnames=false,linktocpage=true]{hyperref}
\hypersetup{colorlinks=true,linkcolor=blue,anchorcolor=blue,citecolor=blue,filecolor=blue,urlcolor=black,bookmarksnumbered=true,
pdfview=FitB
}

\begin{document}

\begin{frontmatter}



\title{Relativistic $^3$He light-front wave function}

\author{V. A. Karmanov\corref{spe}} 
\address{Lebedev Physical Institute,
  Leninsky Prospekt 53, 
119991, Moscow, Russia}
\ead{karmanovva@lebedev.ru}

\author{Zhimin Zhu} 
\ead{zhuzhimin@impcas.ac.cn}

\author{Ziqi Zhang} 
\ead{zhangziqi@impcas.ac.cn}

\author{Kaiyu Fu} 
\ead{kaiyufu94@gmail.com}
\address{Institute of Modern Physics, Chinese Academy of Sciences, \\ No. 509 Nanchang road, 
Lanzhou, 730000, Gansu, China}


\cortext[spe]{Speaker and corresponding author.}


\begin{abstract}
The relativistic light-front (LF) wave function of $^3$He is determined by the three-body LF equation for the Faddeev components in the momentum space.  As an interaction, we take the one-meson exchange kernels, without the potential approximation. 
Within the explicitly covariant formulation of LF dynamics, we calculate the full relativistic $^3$He LF wave function, comprising 32 spin-isospin components.
In the non-relativistic domain, five of these components dominate and closely resemble their non-relativistic counterparts.
Relativistic effects manifest themselves in deviations in relativistic domain of these components from the non-relativistic ones and in appearance of new components. 
\end{abstract}




\end{frontmatter}



\section{Introduction}\label{intro}

One of the important research directions in nuclear structure theory is to investigate the nuclear wave function at high relative nucleon momenta. 
In this regime, the light front dynamics (LFD) has emerged as a robust framework for describing relativistic few-body bound systems, e.g., baryons in quark models and light nuclei composed of high-momentum nucleons. 
However, many existing studies simplify the wave function model by overlooking its complex spin structure and the unique variable dependencies intrinsic to LFD. 
For example, while the non-relativistic deuteron wave function comprises only two components (the S- and D-waves), its relativistic LFD counterpart contains six spin components. 
Each component depends on both the transverse momentum magnitude $|\vec{k}_\perp|$ and the momentum fraction $x$. 
By employing the explicitly covariant LFD framework on the space-time surface $\omega\cd x = 0$ with $\omega^2 = 0$~(see for review \cite{Carbonell:1998rj}),  the deuteron LF wave function, based on one-boson exchange interaction, was successfully calculated \cite{Carbonell:1995yi}.
It allowed to achieve the accurate predictions  of the deuteron electromagnetic form factors  \cite{Carbonell:1999},  in good agreement with the subsequent experimental data \cite{Abbott:2000}.

Recent advances in computational power increased interest in relativistic three-fermion systems.
Thus, the papers~\cite{Polyzou1,Polyzou2,Polyzou3} are devoted  to Poincar\'e invariant treatment of the bound and scattering three-body states.
In the work \cite{Siqi} the nucleon wave function, incorporating not only three-quark Fock sector but also $3q$+gluon and $3q$+$\bar{q}q$, was calculated.
The present work provides the first solution for the relativistic wave function of $^3$He in the explicitly covariant LFD framework. In contrast to \cite{Siqi}, we decompose the wave function in the spin-isospin basis with the scalar coefficients (spin-isospin components).
This wave function involves 32 spin-isospin components, each of which depends on five independent variables, presenting technical challenges that have long remained unresolved. 
Our study focuses on constructing the wave function basis (Sec. \ref{spin-iso}), formulating the system of equations for the spin components (Sec. \ref{equation}),  and presenting the numerical results (Sec. \ref{num}).

\section{Spin-isospin structures and three-fermion wave functions}\label{spin-iso}

$^3$He is a bound state with the total angular momentum $J={1}/{2}$, the total isospin $1/2$, and the positive parity. 
The wave function depends on the spin-isospin projections of the nucleons and the nucleus.
The number of spin projections of three nucleons and $^3$He nucleus corresponds to $2\times2\times2\times2 = 16$ spin structures for the isospin $t=0$ of the nucleon pair. 
There are another 16 spin structures for the pair isospin $t=1$, resulting in the total 32 spin-isospin scalar functions. 
In contrast to the two-body system, parity conservation does not reduce the number of spin components of the $n$-body LF wave function for $n\ge 3$ \cite{Karmanov:1998jp}.  Each component in LFD depends on the transverse momenta $\vec{R}_{1\perp},\,\vec{R}_{2\perp},\,\vec{R}_{3\perp}$ forming scalars and on the momentum fractions $x_1,\, x_2,\, x_3$.
From the constraints $\vec{R}_{1\perp}+\vec{R}_{2\perp}+\vec{R}_{3\perp}=0$ and $x_1+x_2+x_3=1$, we can exclude, say, the variable indexed 3. 
Consequently,
each spin-isospin component depends on five scalars $\{|\vec{R}_{1\perp}|, |\vec{R}_{2\perp}|, \vec{R}_{1\perp}\cd\vec{R}_{2\perp},x_1,x_2\}$.

To ensure proper permutation properties, the full three-body wave function of $^3$He is represented as a sum of three Faddeev components,
\begin{align}
    \Psi(1,2,3)= \Phi_{12}(1,2,3) + \Phi_{12}(2,3,1) + \Phi_{12}(3,1,2).\label{eq:full_WF}
\end{align}
If the Faddeev component is antisymmetric under permutation of the first pair of its arguments $ \Phi_{12}(1,2,3)=  -\Phi_{12}(2,1,3)$, then the full wave function (\ref{eq:full_WF}) is antisymmetric  under permutation of any pair.
Each Faddeev component is a superposition of two isospin states $\xi_{\tau_1\tau_2\tau_3}^{(t)\tau}$ with the pair isospins $t = 0$ and $t = 1$,
\begin{align}
  \Phi_{12}(1,2,3) = \Phi_{12,\sigma_1\sigma_2\sigma_3}^{(0)\sigma}(1,2,3)\xi_{\tau_1\tau_2\tau_3}^{(0)\tau} + \Phi_{12,\sigma_1\sigma_2\sigma_3}^{(1)\sigma}(1,2,3)\xi_{\tau_1\tau_2\tau_3}^{(1)\tau},
  \label{eq:Faddeev_component}
\end{align}
where $\Phi_{12,\sigma_1,\sigma_2,\sigma_3}^{(t)\sigma}$ are the isospin components with definite pair isospins $t$ (total isospin 1/2) and we omit here the spin projections indices ($\sigma_1$, $\sigma_2$, $\sigma_3$, and $\sigma$) and the isospin projections indices ($\tau_1$, $\tau_2$, $\tau_3$, and $\tau$) of the full Faddeev component $\Phi_{12}(1,2,3)$ for brevity.
The isospin functions are defined  by means of the Clebsch-Gordan coefficients, 
\begin{align}\label{xi01}
  \xi^{(0)\tau}_{\tau_1\tau_2\tau_3}=&C^{00}_{\frac{1}{2}\tau_1\,\frac{1}{2}\tau_2} 
  C^{\frac{1}{2}\tau}_{00\,\frac{1}{2}\tau_3},\\
  \xi^{(1)\tau}_{\tau_1\tau_2\tau_3}=&C^{1\tau_{12}}_{\frac{1}{2}\tau_1\,\frac{1}{2}\tau_2}
  C^{\frac{1}{2}\tau}_{1\tau_{12}\,\frac{1}{2}\tau_3}.
\end{align}

Each isospin component $\Phi^{(0,1)}$ can be decomposed in a spin basis $V_{ij{\sigma_1\sigma_2\sigma_3}}^{\sigma}(1,2,3)$
\begin{align}
    \Phi_{12,\sigma_1\sigma_2\sigma_3}^{(0,1)\sigma}(1,2,3) = \sum_{ij} g_{ij}^{(0,1)}(1,2,3)V_{ij{\sigma_1\sigma_2\sigma_3}}^{\sigma}(1,2,3).\label{eq:gij}
\end{align}
where $g_{ij}^{(0,1)}(1,2,3)$ are the spin-isospin scalar functions.

Sixteen orthogonal basis functions $V_{ij}$ ($i,j=1,2,3,4$) can be constructed as follows:
\begin{equation}\label{Vij}
  V_{ij}=C_{ij} [\bar{u}_{\sigma_1}(k_1)T_iU_c\bar{u}_{\sigma_2}(k_2)]\,[\bar{u}_{\sigma_3}(k_3)S_ju^{\sigma}(p)],
\end{equation}
where $\bar{u}_{\sigma_{1,2,3}}(k_{1,2,3})$ are the conjugated nucleon spinors, and $u^{\sigma}(p)$ is spinor of $^3$He. 
$U_c=\gamma_2\gamma_0$ is the charge conjugation matrix. 
The $4\times 4$ matrices $S_j$ and $T_i$ are defined as,
\begin{eqnarray}\label{wf2}
 &&S_1(3)=\frac{1}{N_{S_1}}\left(2x_3-(m+x_3 M)
\displaystyle{\frac{\hat{\omega}}{\omega\cd p}}\right),
\;
S_2(3)=\frac{1}{N_{S_2}}\frac{m}{\omega\cd p}\hat{\omega},
\nonumber\\
&&S_3(3)=\frac{i}{N_{S_3}}\left(2x_3
-(m-x_3 M)\displaystyle{\frac{\hat{\omega}}{\omega\cd p}}\right)\gamma_5,
\;
S_4(3)=\frac{i}{N_{S_4}}\frac{m}{\omega\cd p}\hat{\omega}\gamma_5,
\end{eqnarray}
and 
\begin{eqnarray}\label{wf4}
&&T_1(1,2,3)=\frac{1}{N_{T_1}}\left(2x(1-x)(1-x_3)-
\displaystyle{\frac{m\hat{\omega}}{\omega\cd p}}
\right)i\gamma_5,
\;
T_2=\frac{1}{N_{T_2}}\displaystyle{\frac{m\hat{\omega}}{\omega\cd p}}i\gamma_5,
\nonumber\\
&&T_3(1,2,3)=\frac{1}{N_{T_3}}\left(2x(1-x)(1-x_3)-(1-2x)
\displaystyle{\frac{m\hat{\omega}}{\omega\cd p}}\right),
\;
T_4=\frac{1}{N_{T_4}}\displaystyle{\frac{m\hat{\omega}}{\omega\cd p}}
\nonumber\\
&&
\end{eqnarray}
where $\hat{\omega}=\omega_{\mu}\gamma^{\mu}$,
$$
x=\frac{x_1}{x_1+x_2}=\frac{x_1}{1-x_3}, \quad
\quad x_i=\frac{\omega\cd k_i}{\omega\cd p},
$$
and $T_1(1,2,3)\equiv T_1(x_1,x_2,x_3)$.
These structures are orthogonal and satisfy the following conditions, determining their normalization coefficients $N$:
\footnotesize
\begin{eqnarray*}
\hspace{-1cm}&&\frac{1}{2}\sum_{\sigma_3\sigma}[\bar{u}_{\sigma_3}(k_3)S_{j} u^{\sigma}(p)]^{\dagger}[\bar{u}_{\sigma_3}(k_3)S_{j'} u^{\sigma}(p)]
=\frac{1}{2}\mathrm{Tr}[\bar{S}_j(\hat{k}_3+m)S_{j'}(\hat{p}+M)]=\delta_{jj'},
\\
\hspace{-1cm}&&\sum_{\sigma_1\sigma_2} [\bar{u}_{\sigma_1}(k_1)T_{i}U_c
\bar{u}_{\sigma_2}(k_2)]^{\dagger}
[\bar{u}_{\sigma_1}(k_1)T_{i'}U_c\bar{u}_{\sigma_2}(k_2)]
=-\mathrm{Tr}[\bar{T}_i(\hat{k}_1+m)T_{i'}(-\hat{k}_2+m)]=\delta_{ii'},
\end{eqnarray*}
\normalsize
where $\bar{S}_j=\gamma_0 S_j^{\dagger}\gamma_0$ and similarly for $\bar{T}_i$. The coefficient $\frac{1}{2}$ in the first equality corresponds to averaging over two spin projections $\sigma=\pm \frac{1}{2}$ of $^3$He.

When sandwiched between spinors, some structures exhibit negative parity while others retain positive parity.
The coefficients $C_{ij}$ ensure the positive parity of all $V_{ij}$ and are defined as
\begin{equation}\label{Cij}
C_{ij}(k_1,k_2)=
\left\{
\begin{array}{ll}
1, & \mbox{if $i=1,2, j=1,2$ or $i=3,4,j=3,4$}  
\\
C_{ps}(k_1,k_2),& \mbox{if $i=1,2,j=3,4$ or $i=3,4,j=1,2$} 
\end{array}
\right.,
\end{equation}
where $C_{ps}$ is the pseudoscalar factor,
\begin{eqnarray}\label{wf6}
  C_{ps}(k_1,k_2)=\frac{\varepsilon^{\mu\nu\rho\gamma}k_{1\mu}k_{2\nu}p_{\rho}\omega_{\gamma}}{|\varepsilon^{\mu\nu\rho\gamma}k_{1\mu}k_{2\nu}p_{\rho}\omega_{\gamma}|}.
\end{eqnarray}
Due to the possibility of constructing the pseudoscalar (\ref{wf6}) from the arguments of the LF wave function starting with the three-body one, which allows to "correct" the parity of the spin components, the parity conservation does not reduce the number of the spin components.  
The components with inappropriate parity (say, negative) are multiplied by $C_{ps}(k_1,k_2)$, thereby acquiring appropriate (positive) parity.

The product of the spin and isospin functions can be either symmetric or antisymmetric under permutation of the nucleon pair. 
For example, the spin wave function with spin-1 pair is symmetric, the $t=1$ isospin wave function is also symmetric, so their product remains symmetric. 
In this case, the antisymmetry of the Faddeev component is achieved by antisymmetry of the corresponding scalar momentum-dependent function: 
$g_{ij}^{(1)}(2,1,3)=-g_{ij}^{(1)}(1,2,3)$.

The spin basis $V_{ij}$ is advantageous due to its simplicity and orthonormality, which facilitates equations for the wave function and their solving . However, in the non-relativistic limit, it is not reduced to the conventional basis  \cite{hajduc:1980,baru:2003} used for constructing the non-relativistic wave functions. For example, the simplest term -- corresponding to pair spin 0 and pair isospin 1 -- represented in this basis reads 
$
\tilde{\chi}_1\xi^{(1)\tau}_{\tau_1\tau_2\tau_3}\psi^{(1)}(k,q)
$,
where the spin component $\tilde{\chi}_1$ is expressed via the Clebsch-Gordan coefficients:
\begin{equation} \label{chi1}
  \tilde{\chi}_{1,\sigma_1\sigma_2\sigma_3}^{\sigma}=C^{00}_{\frac{1}{2}\sigma_1\,\frac{1}{2}\sigma_2}
C^{\frac{1}{2}\sigma}_{00\,\frac{1}{2}\sigma_3}Y_{00}Y_{00},
\end{equation}
and $\psi^{(1)}(k,q)$ depends on the modules of the Jacobi momenta  $k$ and $q$.
Other four terms correspond to the pair spins and isospins equal to 1 and 0 and contain  the spherical functions with the angular momentum 2 in addition to $Y_{00}Y_{00}$. 
We replace $\tilde{\chi}_{1,\sigma_1\sigma_2\sigma_3}^{\sigma}$ by its relativistic generalization
\begin{equation}\label{chi1r}
\chi_1=\frac{i}{4\pi\sqrt{2}}c\,[\bar{u}(k_1)\Pi_+\gamma_5U_c\bar{u}(k_2)]\;
[\bar{u}(k_3)\Pi_+u(p)],
\end{equation}
where $\Pi_{\pm}=\frac{{\cal M}\pm \hat{\cal P}}{2\cal{M}}$, ${\cal P}=k_1+k_2+k_3$, and ${\cal M}^2={\cal P}^2$. In the reference frame where $\vec{\cal P}=0$, the function (\ref{chi1r}) (with appropriate coefficient $c$) turns into (\ref{chi1}).

Similarly to (\ref{chi1r}) we construct another 15 new basis functions $\chi_n$  such that the decomposition (\ref{eq:gij}) takes the form:
\begin{align}
 \Phi_{12,\sigma_1\sigma_2\sigma_3}^{(0,1)\sigma}(1,2,3) = \sum_{n = 1}^{16}\chi_{n,\sigma_1\sigma_2\sigma_3}^{\sigma}(1,2,3) \psi^{(0,1)}_n(1,2,3),\label{eq:psin}
\end{align}
These functions are not orthogonal and calculations using them result in rather cumbersome expressions for the kernel matrix. 
Therefore, to solve the equation, we use the basis $V_{ij}$, Eq. (\ref{Vij}), and find the components $g_{ij}$. 
By expressing the basis $\chi_{n,\sigma_1\sigma_2\sigma_3}^{\sigma}(1,2,3)$ in terms of $V_{ij}$, we establish the relation between $\psi^{(0,1)}_n(1,2,3)$ and $g_{ij}$.
The final results will be represented in terms of the relativistic functions $\psi^{(0,1)}_n(1,2,3)$. In the non-relativistic limit, five of them turn into the non-relativistic functions tabulated in \cite{baru:2003}.
Comparing the relativistic solutions for $\psi^{(0,1)}_n(1,2,3)$ with the five non-relativistic ones  \cite{baru:2003}, we will see the relativistic effects in the wave function. 
They also manifest themselves in appearance of other spin-isospin components, absent in the non-relativistic wave function.

\section{Pair interaction and three-body LFD equation}\label{equation}

The three-body  equation with pair interactions can be written symbolically as
\begin{align}
    (H_0-M^2) \Psi = -\Psi \cdot K_{12} -  \Psi \cdot K_{23} -  \Psi \cdot K_{31}. \label{eq:Faddeev_eq}
\end{align}
Substituting the full three-body wave function defined in Eq.~(\ref{eq:full_WF}) into the Faddeev equation in Eq.~(\ref{eq:Faddeev_eq}), and taking into account the permutation properties of kernels and wave functions, we obtain the explicit form of the three-body wave function equation,
\begin{align}
    ({\cal M}^2-M^2)\Phi_{12}(1,2,3)
    =&\sum_{\sigma^\prime_1,\sigma^\prime_2;\tau_1^\prime,\tau^\prime_2}\int \frac{\mathrm{d}^2R^\prime_{2\perp}\mathrm{d}x^\prime_2}{(2\pi)^32x^\prime_1x^\prime_2}
    {\cal K}_{12}(1^\prime,2^{\prime};1,2)\nonumber\\
    \times&\Big[
    \Phi_{12}(1^\prime,2^\prime,3)
    +\Phi_{12}(2^\prime,3,1^\prime) 
    +\Phi_{12}(3,1^\prime,2^\prime)
    \Big]
    ,
    \label{eq:LFD}
\end{align}
where the spin and isospin indices, as well as the explicit arguments of the Faddeev components and their corresponding kernels, are omitted for simplicity.
Here, $x'_1=1-x'_2-x_3$, $M$ is the bound state mass, and the three-body kinetic energy reads,
\begin{equation}\label{Mcal}
   {\cal M}^2 = \frac{\vec{R}_{1\perp}^2 + m^2}{x_1} +  \frac{\vec{R}_{2\perp}^2 + m^2}{x_2} + \frac{\vec{R}_{3\perp}^2 + m^2}{x_3},
\end{equation}
where $m$ is the nucleon mass. The kernel $\mathcal{K}_{12}$ corresponds to the one-boson exchange  between the particle pair 12. The exchanged bosons include pseudoscalar ($\pi$, $\eta$), scalar ($\delta$, $\sigma_0$, $\sigma_1$) and vector ($\omega$, $\rho$) mesons. We take the same parameters (masses, coupling constants) which were used for calculating the LF deuteron wave function (table 2 in \protect{\cite{Carbonell:1995yi}} taken from \protect{\cite{Machleidt:1987hj}}, table 5).

In the framework of explicitly covariant LFD~\cite{Carbonell:1998rj}, the kernel has two contributions corresponding to the two time-ordered diagrams (in the LF time).
Applying the rules of the explicitly covariant LFD~\cite{Carbonell:1998rj},
we obtain the kernel expressions for the scalar and pseudoscalar couplings,
\begin{equation}\label{calK}
    {\cal K}_{12,\sigma_2\sigma_1}^{\sigma'_2\sigma'_1}(1^\prime,2^\prime;1,2)
    =\Pi_{12}(Q^2)F^2(Q^2) \Bigl[\bar{u}^{\sigma_1}(k_1)O_1 u^{\sigma'_1}(k'_1)\Bigr]\,
     \Bigl[\bar{u}^{\sigma_2}(k_2)O_2 u^{\sigma'_2}(k'_2)\Bigr],
\end{equation}
where $\Pi_{12}$ is the scalar part of the propagator,
\begin{equation}\label{Pi12}
    \Pi_{12}(Q^2)=\frac{1}{\mu^2+Q^2}.
\end{equation}
Here, $\mu$ is the meson mass, and the momentum transfer squared $Q^2$ is 
\begin{eqnarray}
Q^2=
    \left\{
    \begin{array}{ll}
    \Bigl({\cal M}^2-M^2\Bigr)(x_1-x'_1)+\frac{m^2(x_1-x'_1)^2+(x_1\vec{R'}_{1\perp}-x'_1\vec{R}_{1\perp})^2}{x_1x'_1},&
    \mbox{if  $x_1-x'_1\geq 0$}
    \\
    \Bigl({{\cal M}'}^2-M^2\Bigr)(x'_1-x_1)+\frac{m^2(x_1-x'_1)^2+(x_1\vec{R'}_{1\perp}-x'_1\vec{R}_{1\perp})^2}{x_1x'_1},&
    \mbox{if  $x_1-x'_1\leq 0$}
    \end{array}
    \right.
,
\end{eqnarray}
where ${\cal M}^2$ is defined in (\ref{Mcal}), and, similarly, 
\begin{equation}\label{Mcalp}
  {\cal M}^{\prime2} = \frac{\vec{R}_{1\perp}^{\prime2} + m^2}{x_1^\prime} + \frac{\vec{R}_{2\perp}^{\prime2} + m^2}{x_2^\prime} + \frac{\vec{R}_{3\perp}^{2} + m^2}{x_3}.
\end{equation}
For scalar exchange, $O_1=O_2=g_s$. For pseudoscalar exchange, $O_1=O_2=i\gamma_5 g_{ps}$. 
When exchanged meson has the isospin $T=1$, an additional factor $\vec{\tau}_1\cd\vec{\tau}_2$ appears. This isospin conserving factor obtains the value $-3$ in the pair $t=0$ isotopic state and 1  in the pair $t=1$ isotopic state.
For the meson isospin $T=0$, this factor is replaced by 1.

For regularization, each vertex is multiplied by the form factor,
\begin{equation}\label{BFF}
    F(Q^2) =\left(\frac{\Lambda^2-\mu^2}{\Lambda^2+ Q^2}\right)^n,
\end{equation}
where $\Lambda$ and $n$ are parameters whose values depend on the coupling \cite{Machleidt:1987hj}.

The kernel for the vector coupling is calculated similarly. Its explicit form can be found in  \protect{\cite{Mangin-Brinet:2003vmv}.}

Substituting the Faddeev components from Eq.~(\ref{eq:Faddeev_component}) into the equation~(\ref{eq:LFD}), and multiplying both parts by 
$V_{ij}^{\dagger}\xi^{(0,1)\tau}_{\tau_1\tau_2\tau_3}$, 
we obtain the integral equation for the scalar functions $g^{(t)}_{ij}(1,2,3)$:
\begin{align}
&(\mathcal{M}^2-M^2)g^{(t)}_{ij}(1,2,3)
\nonumber\\
=&
 \sum_{i^\prime j^\prime t^\prime} \int \frac{\mathrm{d}^2R^\prime_{2\perp}\mathrm{d}x^\prime_2}{(2\pi)^32x^\prime_1x^\prime_2} g^{(t^\prime)}_{i^\prime j^\prime}(1^\prime,2^\prime,3) W_{ijt}^{i^\prime j^\prime t^\prime}\mbox{\tiny (12)}(1^\prime,1;2^\prime,2;3) 
\nonumber\\
+&
  \sum_{i^\prime j^\prime t^\prime} \int \frac{\mathrm{d}^2R^\prime_{2\perp}\mathrm{d}x^\prime_2}{(2\pi)^32x^\prime_1x^\prime_2} g^{(t^\prime)}_{i^\prime j^\prime}(3,1^\prime,2^\prime)W_{ijt}^{i^\prime j^\prime t^\prime}\mbox{\tiny (31)}(1^\prime,1;2^\prime,2;3) 
\nonumber\\
+&
  \sum_{i^\prime j^\prime t^\prime} \int \frac{\mathrm{d}^2R^\prime_{2\perp}\mathrm{d}x^\prime_2}{(2\pi)^32x^\prime_1x^\prime_2} g^{(t^\prime)}_{i^\prime j^\prime}(2^\prime,3,1^\prime)W_{ijt}^{i^\prime j^\prime t^\prime}\mbox{\tiny (23)}(1^\prime,1;2^\prime,2;3) , \label{eq:LFequation}
\end{align}
where $W_{ijt}^{i^\prime j^\prime t^\prime}\mbox{\tiny (12)}$, $W_{ijt}^{i^\prime j^\prime t^\prime}\mbox{\tiny (31)}$, and $W_{ijt}^{i^\prime j^\prime t^\prime}\mbox{\tiny (23)}$ are the products of the kernels $\mathcal{K}_{12}$ and the permuted spin-isospin structures $V_{ij}\xi^{(t)}$.  They are expressed via traces.

\section{Numerical results}\label{num}
Due to the excessive number of degrees of freedom in the integral equation (\ref{eq:LFequation}),  direct numerical solutions face significant challenges.
To overcome this difficulty, we implement an iterative approximation scheme, which was successfully used in   \protect{\cite{Carbonell:1995yi}} for finding the deuteron LF wave function.
Starting with the non-relativistic solution \cite{baru:2003}, we found the components $g_{\text{nr},ij}^{(0,1)}$, substituted them into the r.h.s. of Eq.~(\ref{eq:LFequation}) and, calculated the integral in r.h.s. of this equation. 
This first iteration yields an approximation of the relativistic scalar functions, $g_{\text{r},ij}^{(0,1)}$. 
Subsequently, we obtain the approximate relativistic wave functions $\psi_{\text{r},n}^{(0,1)}$, determining the LF wave functions in the form (\protect{\ref{eq:psin}}).

Figure~1
compares the non-relativistic (nr) scalar functions $g_{\text{nr},ij}^{(0,1)}$ with its relativistic (r) counterparts $g_{\text{r},ij}^{(0,1)}$. 
As explained, the $^3$He LF wave function comprises 32 scalar function components.
For clarity, we display the five dominant non-relativistic and four dominant relativistic components corresponding to pair isospin $t= 0,1$. 
In the non-relativistic domain (at small $q\ll m$ and near $x_{12}=1/2$), the dominant relativistic components align closely with their non-relativistic counterparts. However, deviations emerge in the relativistic domain due to relativistic effects. 
\begin{figure}[htbp!]
    \label{fig:gij_fig}
    \begin{center}
    \includegraphics[width=1\textwidth]{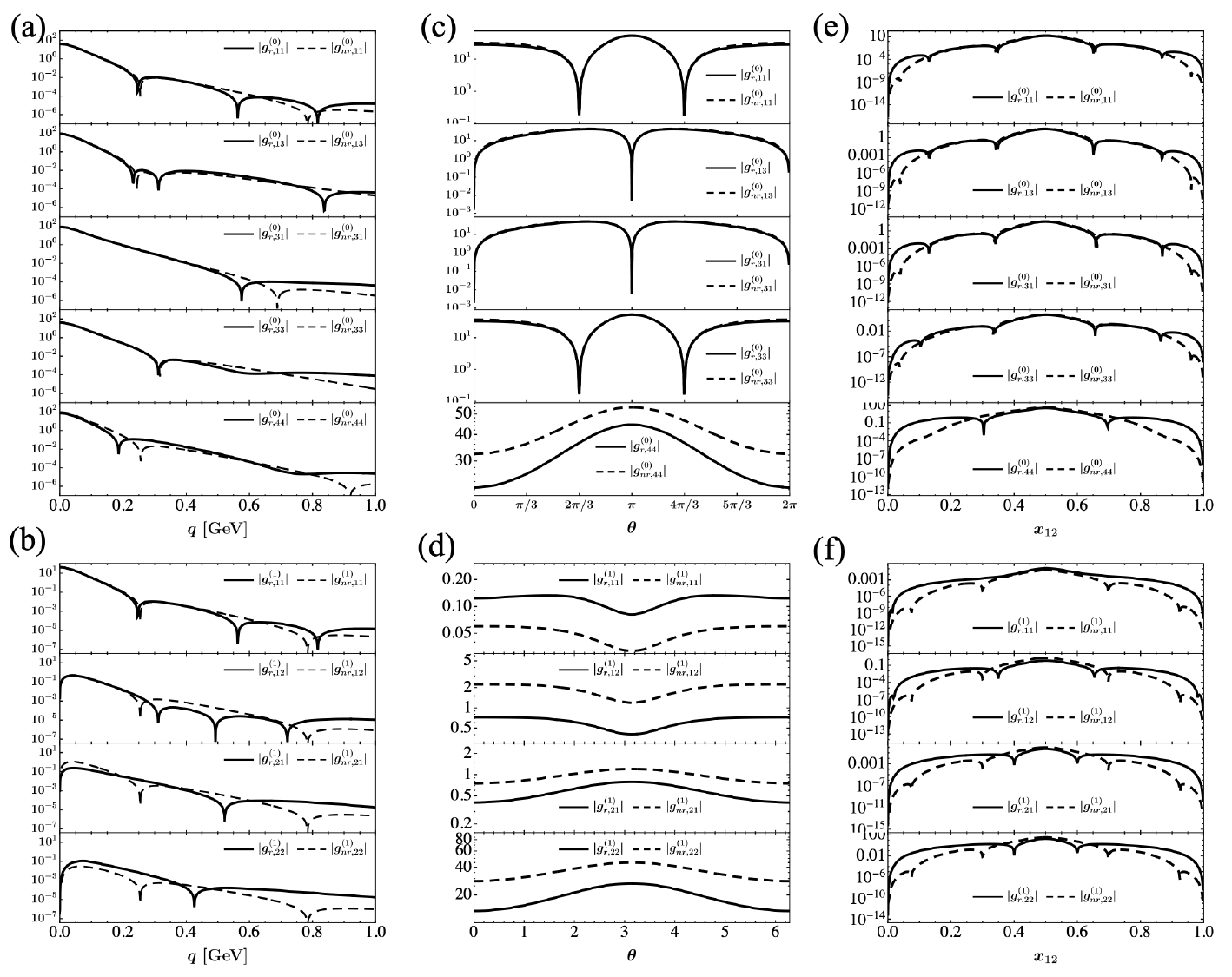}
    \caption{
    (Preliminary results)
    Logarithmic plots of dominant scalar functions with fixed arguments. 
    The solid and dashed lines represent $g_{\text{r},ij}^{(0,1)}$ and $g_{\text{nr},ij}^{(0,1)}$, respectively.
    Panels (a) and (b) represent the functions with the fixed arguments: $x_1 = x_2 = x_3 = 1/3$, $\vec{R}_{1\perp} = (q \cos(\pi/2), q \sin(\pi/2))$, $\vec{R}_{2\perp}=(0.05,0)$, $\vec{R}_{3\perp} = - \vec{R}_{1\perp} - \vec{R}_{2\perp}$; 
    Panels (c) and (d) represent the functions with the fixed arguments: $x_1 = x_2 = x_3 = 1/3$, $\vec{R}_{1\perp} = (0.05 \cos(\theta), 0.05 \sin(\theta))$, $\vec{R}_{2\perp}=(0.05,0)$, $\vec{R}_{3\perp} = - \vec{R}_{1\perp} - \vec{R}_{2\perp}$;
    Panels (e) and (f) represent the functions with the fixed arguments: $x_1 = x_{12}(1-x_3)$, $x_3 = 1/3$, $x_{2} = 1 - x_1 - x_3$, $\vec{R}_{1\perp} = (0.05 \cos(\pi/2), 0.05 \sin(\pi/2))$, $\vec{R}_{2\perp}=(0.05,0)$, $\vec{R}_{3\perp} = - \vec{R}_{1\perp} - \vec{R}_{2\perp}$.
    All transverse momenta are in units of GeV/c.
    }
    \end{center}
\end{figure}

Figure~2
shows the  relativistic spin components $\psi_n$ of $^3$He and five non-relativistic ones (inputs) as functions of $q$. 
Among the 32 components, the relativistic spin components $\psi_{\text{r},1}^{(1)}$ and $\psi_{\text{r},2}^{(0)}$ are the dominant ones, similar to their non-relativistic counterparts $\psi_{\text{nr},1}^{(1)}$ and $\psi_{\mathrm{nr},2}^{(0)}$. 
In the non-relativistic domain, these two components closely resemble their non-relativistic counterparts.
The other three wave functions $\psi_{\text{r},3-5}^{(0)}$ with non-relativistic counterparts have some deviations from their non-relativistic ones in the non-relativistic domain. 
The appearance of other spin components which have no the non-relativistic counterparts is due to the relativistic effects.
The smaller amplitudes of these components indicate that relativistic effects remain relatively weak, which is consistent with intuition.
\begin{figure}[htbp!]
    \begin{center}
    \includegraphics[width=0.45\textwidth]{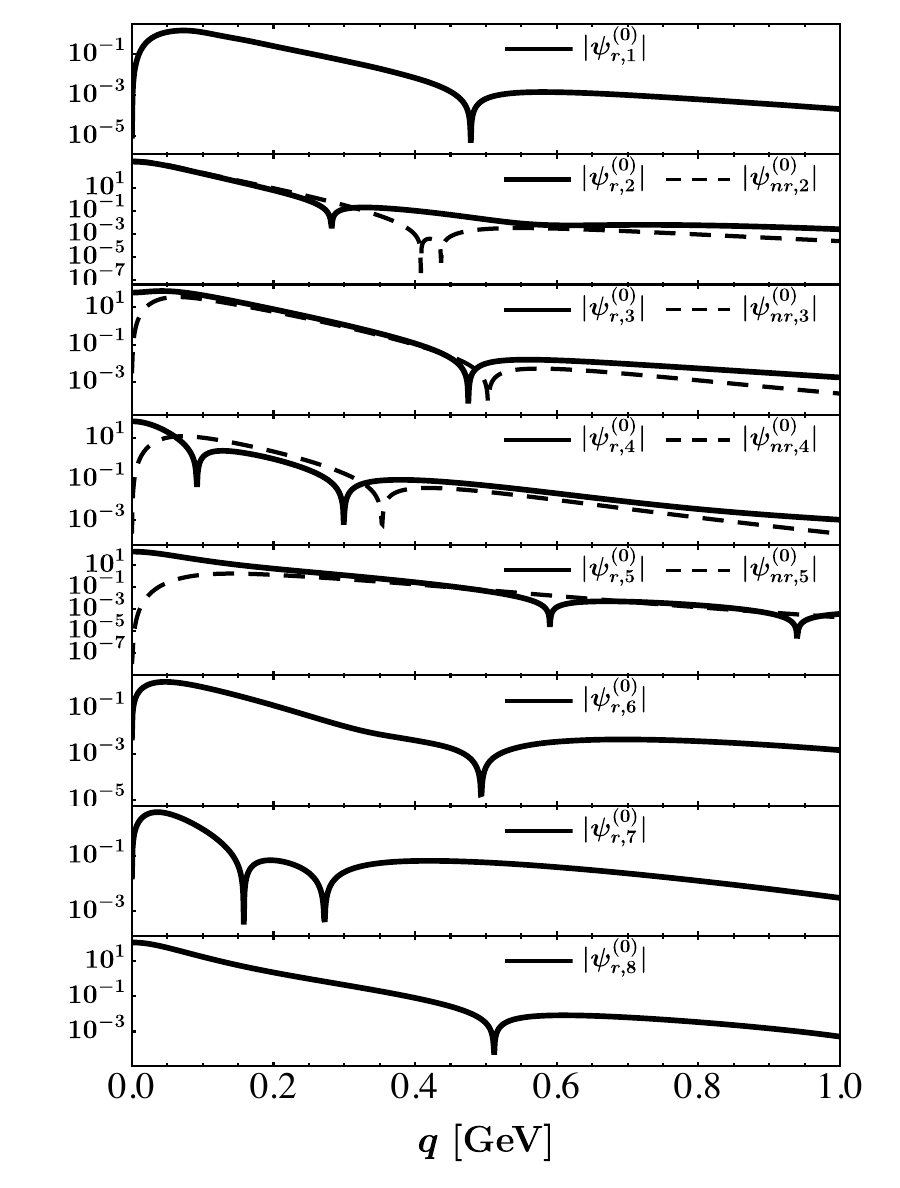} 
    \includegraphics[width=0.45\textwidth]{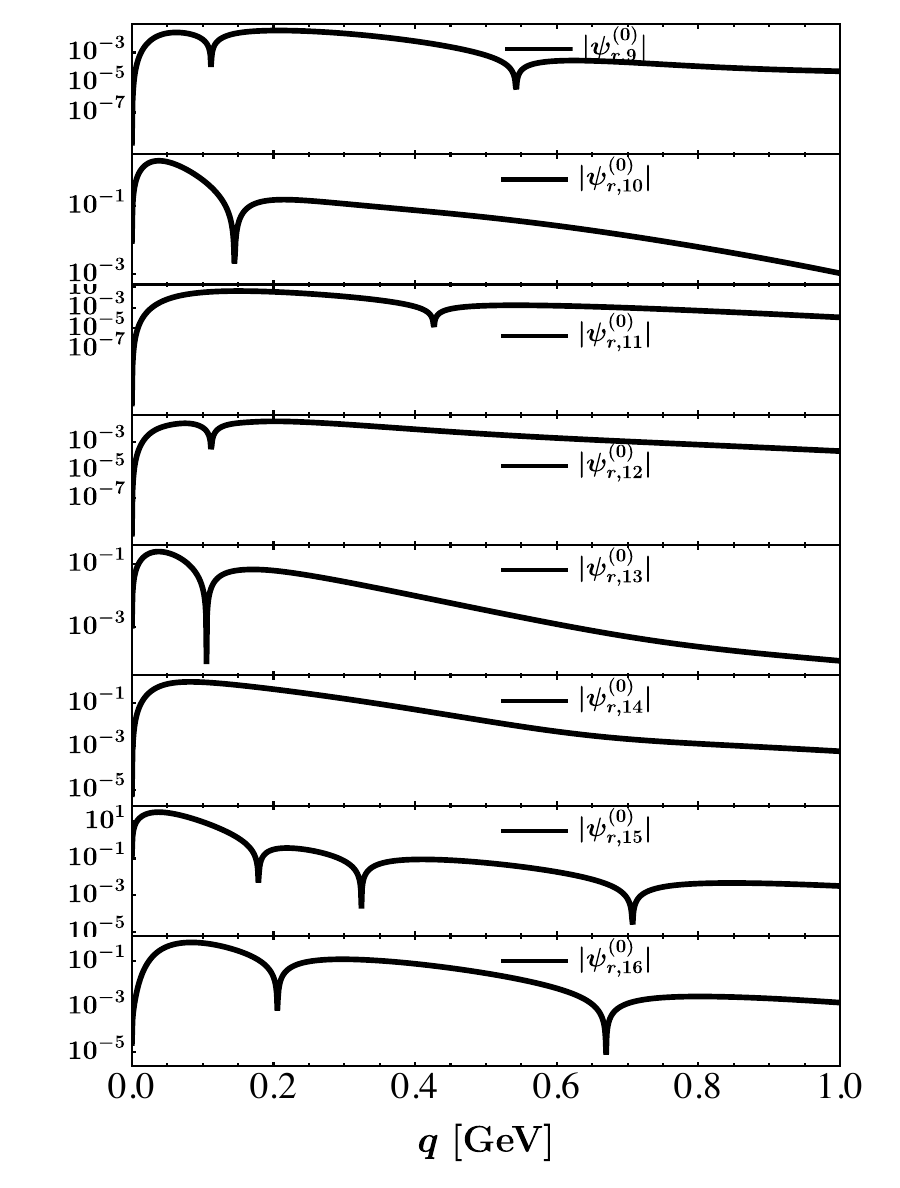} 
    \includegraphics[width=0.45\textwidth]{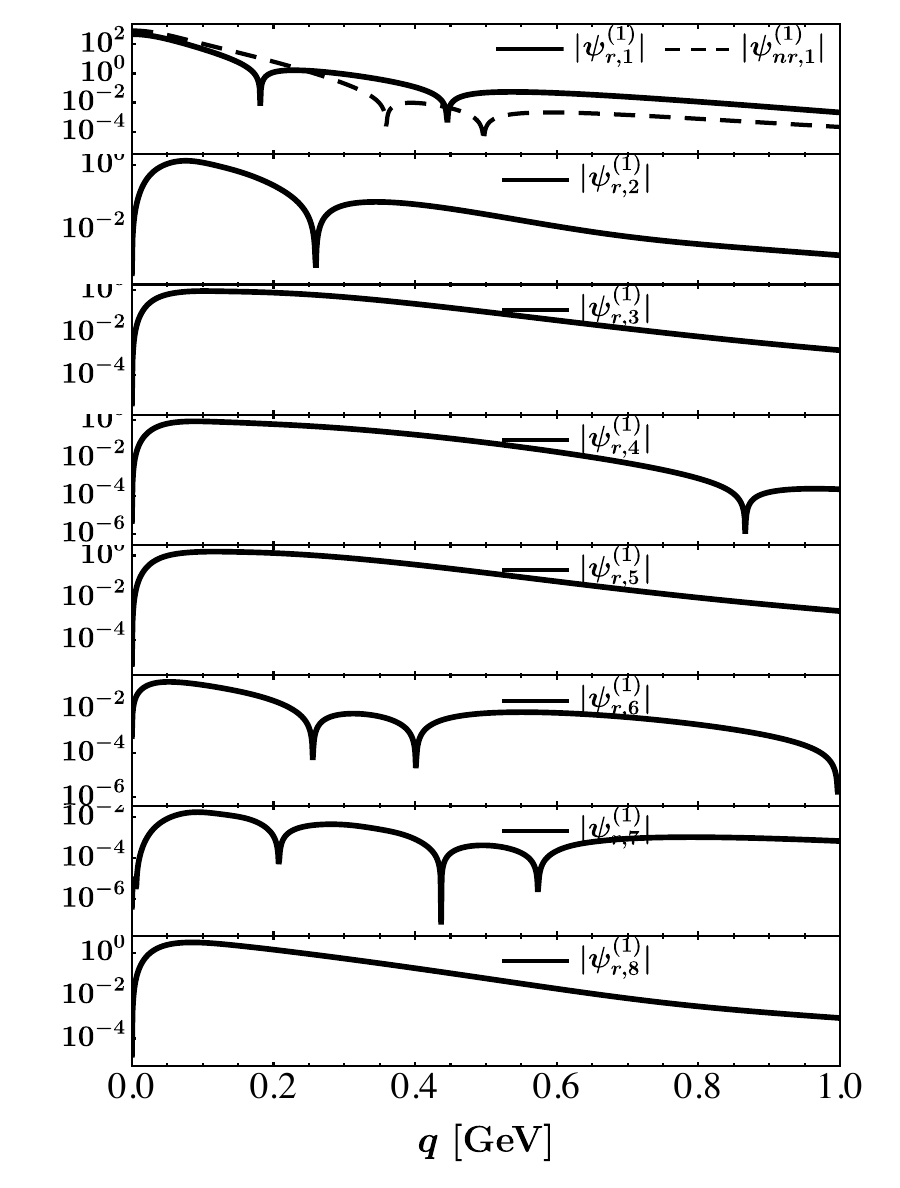} 
    \includegraphics[width=0.45\textwidth]{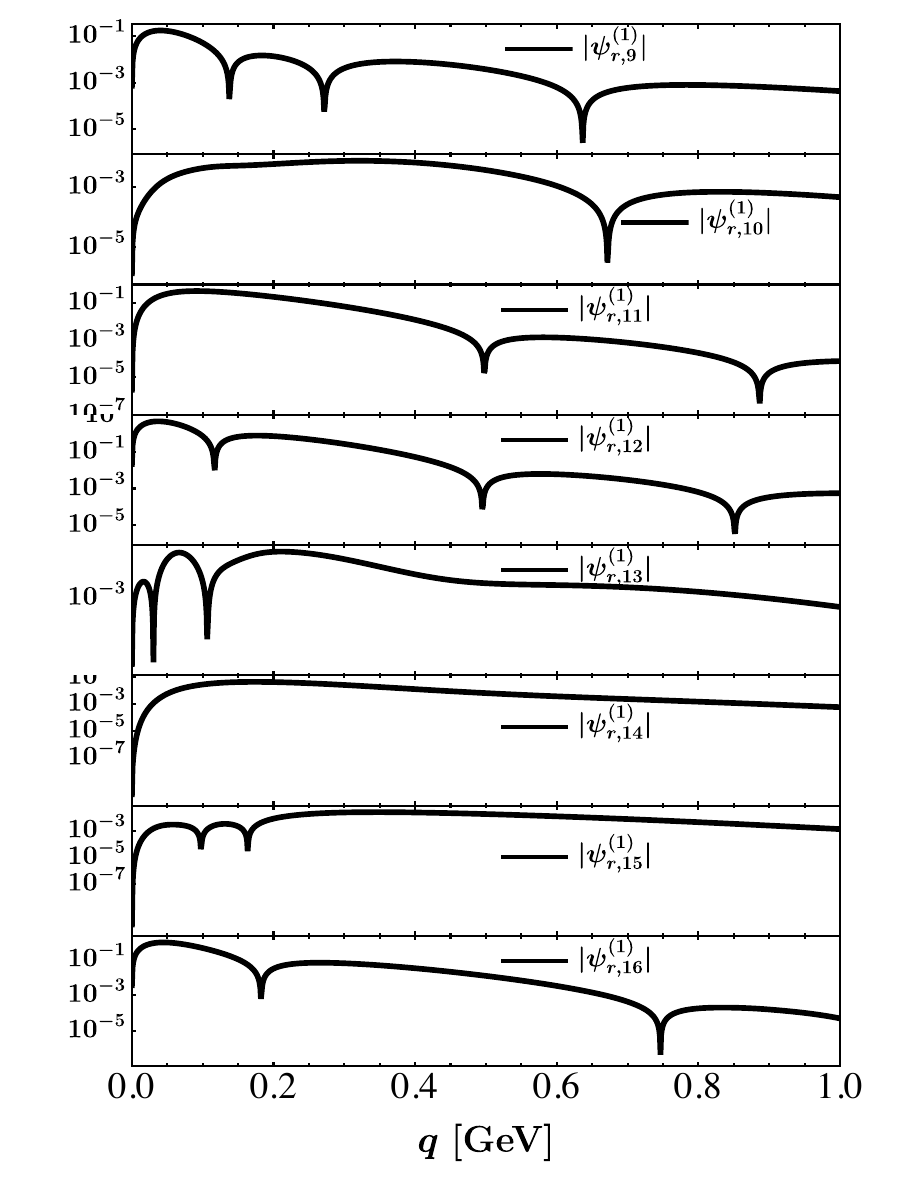} 
    \caption{
    (Preliminary results)
    Logarithmic plots of relativistic wave functions ($\psi_{\text{r},n}^{(0,1)}$, solid lines) and non-relativistic wave functions ($\psi_{\text{nr},n}^{(0,1)}$, dashed lines) under fixed arguments, $x_1 = x_2 = x_3 = 1/3$, $\vec{R}_{1\perp} = (q \cos(\pi/2), q \sin(\pi/2))$, $\vec{R}_{2\perp}=(0.05,0)$, $\vec{R}_{3\perp} = - \vec{R}_{1\perp} - \vec{R}_{2\perp}$.
    All transverse momenta are in units of GeV/c.
    }
    \end{center}
    \label{fig:psi_plot}
\end{figure}

\section{Conclusion}
In this work, we calculate the relativistic LF wave function of $^3$He within a three-body relativistic framework.  The full LF wave functions consist of all 32 spin-isospin components. 
Our results demonstrate a consistent trend between the dominant relativistic wave functions and the non-relativistic ones in the non-relativistic domain. 
Additionally, the amplitudes of the wave functions without non-relativistic counterparts suggest that relativistic effects remain relatively weak in the domain of momenta considered in the present work.
Nevertheless, incorporating all relativistic components is essential for accurate computing the electromagnetic form factors, momentum distributions, and other observables. Neglecting these components in relativistic calculations introduces uncontrolled approximations, which can significantly affect theoretical predictions.
In future work, we aim to compute these observables using the relativistic $^3$He wave functions. Moreover, our approach is general and applicable to any three-fermion,  including the nucleon, and to many-body relativistic systems.  It is "canonical" in the sense that the spinors and the spin structures formed by 
Dirac matrices sandwiched with 
the spinors, which are covariant, that is, they are properly transformed under rotations and the Lorentz transformations, are constructed analytically by Eq. (\ref{Vij}). Whereas, all the dynamics resulted from interaction between the fermions, is included in the invariant functions $g_{ij}^{(t)}$ determining the Faddeev components (\ref{eq:gij}) of the full wave function (\ref{eq:full_WF}). 

Another LF approach \cite{Siqi}, applied to the nucleon wave function, uses decomposition of the  LF wave function 
in harmonic oscillator basis and solves system of equations for the coefficients of this decomposition. The covariant spin structures are not separated. Therefore the transformation properties are hidden in the dependence of these coefficients on the spin projections of the constituents and of the bound state
in different systems of reference, subjected to rotations and the Lorentz transformations. Numerical description of these extra (kinematical) degrees of freedom requires extra supercomputer power which is available. 
Both researches are treating, in the framework of one and the same general approach - LFD, by different technical methods, one and the same problem and can be considered as complementary to each other. 

Future extensions of the explicitly covariant LFD framework will involve applications to  nucleons, to other light nuclei, such as $^4$He, and to the electromagnetic form factors.
\bigskip

\noindent
{\bf Acknowledgement}

VAK was supported by the Chinese Academy of Sciences President's International Fellowship Initiative (PIFI), Grant No. 2023VMA0010, during his visits to Institute of Modern Physics of Chinese Academy of Sciences, where part of this work was fulfilled.
VAK is sincerely grateful to the light-front QCD group in IMP for kind hospitality.

 Z. Zhu is supported by the Natural Science Foundation of Gansu Province China Grant No. 23JRRA571. Z. Zhu and Z. Zhang gratefully acknowledge financial support from China Scholarship Council.
This research is supported by Gansu International Collaboration and Talents Recruitment Base of Particle Physics (2023–2027), by the Senior Scientist Program funded by Gansu Province Grant No. 25RCKA008.

\end{document}